%% file: main.tex
\documentclass[aps, prc,twocolumn, superscriptaddress, showkeys, nofootinbib]{revtex4-1}

\usepackage[margin=1in]{geometry}
\usepackage{amsmath,amssymb,amsthm,bm,mathtools}
\usepackage{physics}
\usepackage{booktabs}
\usepackage{enumitem}
\usepackage{graphicx,color}
\usepackage{amsmath,amssymb,amsfonts}
\usepackage{hyperref}

\def\etas{\mbox{$\eta / {s}$ }}

\usepackage[utf8]{inputenc}

\usepackage{xspace}	% Include xspace

\begin{document}

\title{
Rapidity-even dipolar flow in relativistic heavy-ion collisions
}
%----------------------------------------------------------------------------
\medskip
%----------------------------------------------------------------------------
%----------------------------------------------------------------------------
\author{Niseem~Magdy}
\email{niseem.abdelrahman@tsu.edu}
\affiliation{Department of Physics, Texas Southern University, Houston, TX 77004, USA}
\affiliation{Physics Department, Brookhaven National Laboratory, Upton, New York 11973, USA}
%----------------------------------------------------------------------------

%----------------------------------------------------------------------------
%\date{\today}
%----------------------------------------------------------------------------
\begin{abstract}
%----------------------------------------------------------------------------
Rapidity-even directed flow, $v_{1}^{\rm even}$, provides a sensitive probe of fluctuation-driven dipolar asymmetry in relativistic heavy-ion collisions, but its extraction is complicated by strong first-harmonic non-flow correlations, particularly those associated with global momentum conservation (GMC). In this study, $v_{1}^{\rm even}$ and its multi-particle correlations are investigated in Au+Au collisions at $\sqrt{s_{NN}}=200$\~GeV using the AMPT and HIJING models. An $\eta$-dependent weighting procedure is applied to suppress the leading GMC contribution. The GMC-suppressed HIJING results are strongly reduced for most $v_1$-related observables, while AMPT reproduces the characteristic sign-changing $p_T$ dependence of $v_{1}^{\rm even}$ and exhibits clear sensitivity to the partonic transport setting. The non-normalized mixed-harmonic correlations show a stronger transport dependence, whereas the normalized observables are substantially less sensitive to the absolute final-state response. Comparison with the initial-state eccentricity correlations shows that these normalized observables retain the underlying dipolar, elliptic, and triangular correlation structure after modification by the medium. These results demonstrate that GMC-suppressed rapidity-even dipolar flow correlations provide complementary constraints on initial-state fluctuations and their transformation during the final-state evolution.

%----------------------------------------------------------------------
\end{abstract}
%-------------------------------
%\pacs{25.75.-q, 25.75.Gz, 25.75.Ld}% PACS, the Physics and Astronomy % Classification Scheme.
%-------------------------------
%Use showkeys class option if keyword
%-------------------------------
\maketitle
%-------------------------------
%\linenumbers
%-------------------------------
%-------------------------------
\section{Introduction}
%-------------------------------

Ultra-relativistic heavy-ion collisions at The Relativistic Heavy Ion Collider (RHIC) and The Large Hadron Collider (LHC) create a short-lived deconfined medium of quarks and gluons, the quark-gluon plasma (QGP)~\cite{Shuryak:1978ij, Shuryak:1980tp, Muller:2012zq}, whose collective expansion retains memory of the initial collision geometry. The azimuthal anisotropy of final-state particle production is commonly characterized by the Fourier coefficients $v_n$~\cite{Voloshin:1994mz, Poskanzer:1998yz},
\begin{equation}
\frac{dN}{d\phi}
\propto
1+2\sum_{n=1}^{\infty}v_n\cos n(\phi-\Psi_n),
\label{eq:fourier}
\end{equation}
where $v_n$ and $\Psi_n$ denote the magnitude and event-plane angle of the $n^{\rm th}$-order harmonic. These coefficients provide a quantitative connection between the initial-state geometry, its event-by-event fluctuations, and the transport properties of the QGP~\cite{Teaney:2010vd, Bhalerao:2014xra, Yan:2015jma, Bozek:2020drh, Alver:2008zza, Giacalone:2020byk, Schenke:2014tga, Staig:2010pn, Heinz:2001xi, Hirano:2005xf, Huovinen:2001cy, Hirano:2002ds, Romatschke:2007mq, Luzum:2011mm, Song:2010mg, Qian:2016fpi, Magdy:2017ohf, Magdy:2017kji, Bilandzic:2021voo, Schenke:2011tv, Teaney:2012ke, Gardim:2012yp, Lacey:2013eia, Bhatta:2022dml}.

The elliptic flow harmonic ($v_2$) is largely driven by the average almond-shaped overlap geometry in non-central collisions, whereas higher-order harmonics, particularly the triangular flow harmonic ($v_3$), arise predominantly from fluctuations in the participant density. Such fluctuations generate localized hotspots and irregular pressure gradients that are converted into final-state momentum anisotropies through collective expansion. Within this framework, the rapidity-even component of directed flow, $v_1^{\rm even}$, has emerged as a sensitive probe of the fluctuating initial state~\cite{Teaney:2010vd, Retinskaya:2012ky, Jia:2012gu}.

Directed flow was traditionally associated with the sideward deflection of matter in the reaction plane. This conventional component is rapidity odd, $v_1^{\rm odd}(\eta)=-v_1^{\rm odd}(-\eta)$, and is connected to the initial tilt of the fireball, baryon stopping, and the equation of state~\cite{Snellings:1999bt, Zabrodin:2000xc, Borghini:2002vp, STAR:2003xyj, PHOBOS:2005ylx}. In contrast, the rapidity-even component, $v_1^{\rm even}(\eta)=v_1^{\rm even}(-\eta)$, is generated by dipolar fluctuations of the initial density profile. It therefore reflects the final-state response to a fluctuation-induced dipole asymmetry rather than a remnant of global sideward motion~\cite{Teaney:2010vd, Luzum:2010fb, Retinskaya:2012ky, Jia:2012gu}.

The relevant initial-state quantity is the dipolar eccentricity,
\begin{equation}
\varepsilon_1 e^{i\Phi_1}
=
-
\frac{\int r^3 e^{i\phi}\rho(r,\phi)\,d^2r}
{\int r^3 \rho(r,\phi)\,d^2r},
\label{eq:eps1}
\end{equation}
where $\rho(r,\phi)$ is the initial transverse density distribution and $\Phi_1$ is the dipole participant-plane angle~\cite{Teaney:2010vd}. The $r^3$ weighting suppresses the trivial center-of-mass displacement and emphasizes peripheral density fluctuations, making $\varepsilon_1$ especially sensitive to hotspots, surface structure, nuclear substructure, and neutron-skin effects~\cite{Magdy:2023fsp, Shafi:2025feq}.

The final-state anisotropic flow harmonics can be written schematically as a response to the corresponding initial-state eccentricity, with possible nonlinear mode-coupling contributions~\cite{STAR:2019zaf, Liu:2018hjh, STAR:2022vkx}:
\begin{eqnarray}
v^2_n &=&  ( v^{L}_{n} )^2 + ( v^{NL}_{n} )^2  \nonumber \\ 
v^2_n &\sim& (\kappa_n^L \varepsilon_n^L)^2 + (\kappa_n^{NL}\varepsilon_n^{NL})^2,
\label{response}
\end{eqnarray}
where \(v_n^{\rm L}\) and \(v_n^{\rm NL}\) denote the linear and nonlinear contributions to the final-state flow harmonic \(v_n\), respectively. The coefficients \(\kappa_n^{\rm L}\) and \(\kappa_n^{\rm NL}\) represent the corresponding linear and nonlinear medium-response coefficients. Similarly, \(\varepsilon_n^{\rm L}\) and \(\varepsilon_n^{\rm NL}\) denote the linear and nonlinear components of the initial-state eccentricity that drive the corresponding flow response. For $n=2$ and $n=3$ in large A+A collisions, the linear response to $\varepsilon_2$ and $\varepsilon_3$ is expected to dominate.
% to a good approximation, although nonlinear contributions can still be present. 
The dipolar harmonic is qualitatively different because its $p_T$ dependence is constrained by total transverse-momentum conservation. Consequently, $v_1^{\rm even}$ should not be assumed to follow the same simple linear-response scaling as $v_2$ and $v_3$, see Appendix~\ref{app:kappa}.

The dipole response differs qualitatively from higher harmonics because the total transverse momentum of the system must vanish. This imposes the approximate constraint
\begin{equation}
%\int dp_T\,p_T\,v_1^{\rm even}(p_T)\simeq 0,
 \left\langle p_T v_1^{\rm even}(p_T)\right\rangle \simeq 0,
\label{eq:ptconstraint}
\end{equation}
which leads to a sign change of $v_1^{\rm even}(p_T)$: low- and high-$p_T$ particles preferentially contribute with opposite signs. This behavior has been predicted in hydrodynamic calculations and observed in experimental and transport-model studies~\cite{Retinskaya:2012ky, STAR:2018gji, Jia:2012gu}. Consequently, $v_1^{\rm even}$ is sensitive not only to local pressure gradients and initial dipolar geometry, but also to global recoil constraints and the efficiency of the collective medium response~\cite{Retinskaya:2012ky}.

A central challenge in measuring $v_1^{\rm even}$ is the large non-flow background generated by global momentum conservation (GMC). Momentum conservation induces an anti-correlation between particles and contributes directly to the first harmonic of two-particle azimuthal correlations. The flow coefficient can be approximated as
\begin{equation}
v_{1,1}(p_T^a,p_T^b)
=
v_1^{\rm even}(p_T^a)v_1^{\rm even}(p_T^b)
-
c\,p_T^a p_T^b,
\label{eq:v11}
\end{equation}
where the second term represents the GMC contribution~\cite{Borghini:2002mv,Retinskaya:2012ky,Jia:2012gu}. Since this background can be comparable to or larger than the collective signal, its suppression is essential for a reliable extraction of $v_1^{\rm even}$. Moreover, transport studies indicate that the effective GMC contribution can depend on pseudorapidity, transverse momentum, and event activity~\cite{Jia:2012gu}. Thus, a simple pseudorapidity gap or standard $\eta$-independent correction is generally insufficient, motivating a more differential treatment based on $\eta$-dependent weighting.

The fluctuation-driven origin of $v_1^{\rm even}$ further motivates the study of its fluctuations and correlations with other harmonics. Correlations between $v_1^{\rm even}$ and $v_3$ can test whether common hotspots generate both dipolar and triangular structures, while correlations between $v_1^{\rm even}$ and $v_2$ probe how the global elliptic geometry biases the dipolar perturbation. Multi-particle observables, including symmetric and asymmetric correlations and their normalized counterparts~\cite{Magdy:2024ooh}, provide a natural framework for separating magnitude correlations, event-plane correlations, and residual non-flow effects after GMC suppression.

These observables are especially relevant for systems with nontrivial nuclear structure. Clustered configurations, neutron skins, and surface diffuseness can modify the local density profile and may affect the dipolar eccentricity more strongly than conventional harmonics. Recent studies of isobaric and light-ion collisions suggest that rapidity-even dipolar flow and its correlations can provide sensitivity to nuclear geometry beyond that accessible from $v_2$ and $v_3$ alone~\cite{Jia:2021tzt, Magdy:2023fsp, Shafi:2025feq}. This motivates a systematic investigation of $v_1^{\rm even}$, its fluctuations, and its correlations with elliptic and triangular flow.

In this work, rapidity-even dipolar flow is investigated in Au+Au collisions at $\sqrt{s_{NN}}=200$ GeV using the AMPT and HIJING models. An $\eta$-dependent weighting procedure is employed to suppress the GMC-induced backgrounds, and the resulting $v_1^{\rm even}$ is studied as a function of transverse momentum and centrality. Multi-particle correlations involving $v_1$, $v_2$, and $v_3$ are then examined to quantify the interplay between dipolar fluctuations, elliptic geometry, and triangular flow. AMPT is used to investigate the role of the final-state medium response, while HIJING provides a non-collective reference that includes global momentum conservation, jets/minijets, string fragmentation, and associated few-particle correlations. Together, these studies provide a unified framework for assessing the sensitivity of rapidity-even dipolar flow and its correlations to both initial-state geometry and final-state transport.
%--------------------------------------------------------------------
%--------------------------------------------------------------------

%--------------------------------------------------------------------
%--------------------------------------------------------------------
\section{Methodology}
%--------------------------------------------------------------------
\subsection{Models} \label{sec:2}
%The string-melting version of AMPT is used throughout this work.
%--------------------------------------------------------------------
The present analysis employs events generated with the AMPT (v2.26t9b) ~\cite{Lin:2004en} and HIJING (v1.411)~\cite{Wang:1991hta, Gyulassy:1994ew} models for Au+Au collisions at $\sqrt{s_{NN}} = 200$ GeV. Charged particles within the kinematic range $0.2 < p_T < 4.0$~(GeV/$c$) and pseudorapidity $|\eta| < 1.0$ are used as particles of interest (POI), while charged particles in the range $0.2 < p_T < 2.0$~(GeV/$c$) are used as particles of reference (POR). %The HIJING model, which contains only non-flow correlations, is utilized to estimate non-flow contributions, whereas the AMPT model is used to investigate the sensitivity of the flow correlators to final-state medium effects.

\subsubsection{HIJING}
The HIJING model~\cite{Wang:1991hta, Gyulassy:1994ew} is a Monte Carlo event generator developed to describe parton and particle production in high-energy hadronic and nuclear collisions. Heavy-ion collisions are modeled using Glauber geometry through independent binary nucleon--nucleon interactions. HIJING is widely used to study jet and mini-jet production, together with the associated particle production, in p+p, p+A, and A+A collisions. The model incorporates PYTHIA~\cite{Sjostrand:1986hx} to determine the kinematic properties of scattered partons from hard and semihard processes, while the hadronization stage is treated using the Lund string fragmentation model~\cite{Andersson:1983ia}. In this work, HIJING is used with its default settings.

\subsubsection{AMPT}
The AMPT model has been extensively used to investigate various aspects of relativistic heavy-ion collisions~\cite{Lin:2004en, Ma:2013uqa, Nie:2018xog, Magdy:2020bhd, Magdy:2022cvt}. The model consists of several main components: (i) an initial partonic stage generated by the HIJING model~\cite{Wang:1991hta, Gyulassy:1994ew}, where particle production is governed by the Lund string fragmentation function~\cite{Ferreres-Sole:2018vgo},
\begin{equation}
f(z)\propto z^{-1}(1-z)^a \exp\left(-\frac{b\,m_\perp^2}{z}\right),
\end{equation}
with the parameters $a=0.55$ and $b=0.15$ GeV$^{-2}$ as described in Ref.~\cite{Xu:2011fi}. Here, $z$ denotes the light-cone momentum fraction carried by the produced hadron with transverse mass $m_\perp$ relative to the fragmenting string. (ii) A partonic scattering stage characterized by an effective parton scattering cross section,
%--------------------------------------------------------------------
\begin{eqnarray}\label{eq:21}
\sigma_{\rm parton} &=& \dfrac{9 \pi \alpha^{2}_{s}}{2 \mu^{2}},
\end{eqnarray}
%--------------------------------------------------------------------
where $\mu$ denotes the screening mass and $\alpha_s$ represents the QCD coupling constant, both of which govern the partonic scattering dynamics and the subsequent expansion of the medium in A+A collisions~\cite{Zhang:1997ej}. The AMPT model further includes (iii) hadronization through a quark coalescence mechanism, followed by hadronic rescattering described by a hadronic transport approach~\cite{Li:1995pra}. Within the present AMPT framework, only the initial value of $\eta/s$ at the early stage of the collision can be estimated through an appropriate choice of the partonic scattering cross section $\sigma_{\rm parton}$ by varying $\mu$ and/or $\alpha_s$. The corresponding values 
of $\eta/s$ quoted in Table~\ref{tab:1} should be understood as approximate estimates of the initial partonic shear-viscosity to entropy-density ratio obtained from Eq.~(\ref{eq:22}) under the assumptions of a gas of massless quarks and gluons at $T_i=378$~MeV. They are therefore not treated as precision hydrodynamic values of $\eta/s$, but rather as labels for two AMPT transport settings with different partonic scattering strengths~\cite{Xu:2011fi, Nasim:2016rfv, Solanki:2012ne}:
\begin{eqnarray} \label{eq:22}
 \dfrac{\eta}{s} &=& \dfrac{3 \pi}{40 \alpha^{2}_{s}}  \dfrac{1}{ \left(  9 +  \dfrac{\mu^2}{T_{i}^2} \right)  \ln\left(\dfrac{18 + \mu^2/T_{i}^2}{ \mu^2/T_{i}^2 } \right) - 18}.
\end{eqnarray}
%--------------------------------------------------------------------

%--------------------------------------------------------------------
\begin{table}[h!]
\begin{center}
\caption{The summary of the AMPT sets used in this work.\label{tab:1}}
 \begin{tabular}{|c|c|c|c|c|}
 \hline 
 AMPT-set & $\mu$~($fm^{-1}$) & $\alpha_{s}$ & $\sigma_{\rm parton}~(mb)$ &  \etas     \\
  \hline
  Set-1   &     1.27          &     0.33     &     9.6            &   0.125   \\    
  \hline
  Set-2   &     1.80          &     0.33     &     4.8            &   0.175   \\ 
 \hline
\end{tabular} 
\end{center}
\end{table}
%--------------------------------------------------------------------

This study simulates minimum-bias Au+Au collisions at $\sqrt{s_{\rm NN}}=200$~GeV using AMPT with string-melting and HIJING, generating approximately $6.0\times 10^7$ events for each model configuration. Centrality is defined using cuts on the number of participating nucleons, $N_{\rm part}$. Only statistical uncertainties are shown throughout this work. %The variation between the two AMPT transport settings is used to illustrate model-parameter sensitivity; a full systematic uncertainty associated with model-parameter variations is left for future work.
%--------------------------------------------------------------------

%Au+Au collisions at $\sqrt{s_{\rm NN}}=$~200~GeV.
%

%--------------------------------------------------------------------
\subsection{Observables}~\label{sec:3}
%--------------------------------------------------------------------
\subsubsection{Momentum-conservation suppression}
\label{subsec:gmc_weight}

The measurement of rapidity-even directed flow is strongly affected by GMC, which to leading order is given by Eq.~\ref{eq:v11}~\cite{Danielewicz:1985hn, Borghini:2002mv, Luzum:2010fb, Retinskaya:2012ky, Jia:2012gu}. Since this contribution can be comparable to the genuine dipolar-flow signal, it must be suppressed before interpreting $v_1^{\rm even}$ and its correlations with higher harmonics.
%-----------------------------------------------

%-----------------------------------------------
In this work, the GMC correction is implemented through an $\eta$-dependent weighting procedure~\cite{Luzum:2010fb}. For the first harmonic, the reference flow vector is constructed with a weight designed to make the dipolar vector approximately orthogonal to the transverse-momentum vector in each $\eta$ interval. The corresponding first-harmonic weight is written as:
\begin{equation}
\omega(p_T,\eta) = p_T - w(\eta), \label{eq:w1}
\end{equation}
\begin{equation}
    w(\eta)=
    \frac{\langle p_T^2\rangle_{\eta}}
         {\langle p_T\rangle_{\eta}} .
    \label{eq:weta}
\end{equation}
The present implementation uses the $\eta$-dependent generalization discussed in Refs.~\cite{Luzum:2010fb, Jia:2012ma}, in which $\langle p_T\rangle$ and $\langle p_T^2\rangle$ are evaluated locally in pseudorapidity. For a symmetric collision system and acceptance, $w(\eta)$ and $\omega(p_T,\eta)$ are even under $\eta \rightarrow -\eta$. Consequently, integration over a symmetric pseudorapidity interval suppresses the rapidity-odd directed-flow component while retaining the rapidity-even component. This differs from traditional measurements of rapidity-odd directed flow~\cite{Danielewicz:1985hn}.
The averages entering $w(\eta)$ are evaluated using the same reference-particle interval, $0.2<p_T<2.0$~GeV/$c$, that is used to construct the first-harmonic reference vector. This choice gives:
\begin{equation}
    \left\langle \omega(p_T,\eta)\,p_T \right\rangle_{\eta}
    \simeq 0,
    \label{eq:orthogonality}
\end{equation}
thereby suppressing the leading momentum-conservation contribution 
locally in pseudorapidity. The first-harmonic reference vector is then 
constructed as
\begin{equation}
    Q_1 =
    \sum_{i=1}^{M}
    \omega(p_{T,i},\eta_i)e^{i\phi_i},
    \label{eq:q1}
\end{equation}
where the sum runs over particles of reference. For harmonics 
$n>1$, the standard unweighted flow vectors are used,
\begin{equation}
    Q_n =
    \sum_{i=1}^{M} e^{in\phi_i}.
    \label{eq:qn}
\end{equation}
%-----------------------------------------------
%-----------------------------------------------
\begin{figure}[htb]
\centering
\includegraphics[width=1.0\linewidth]{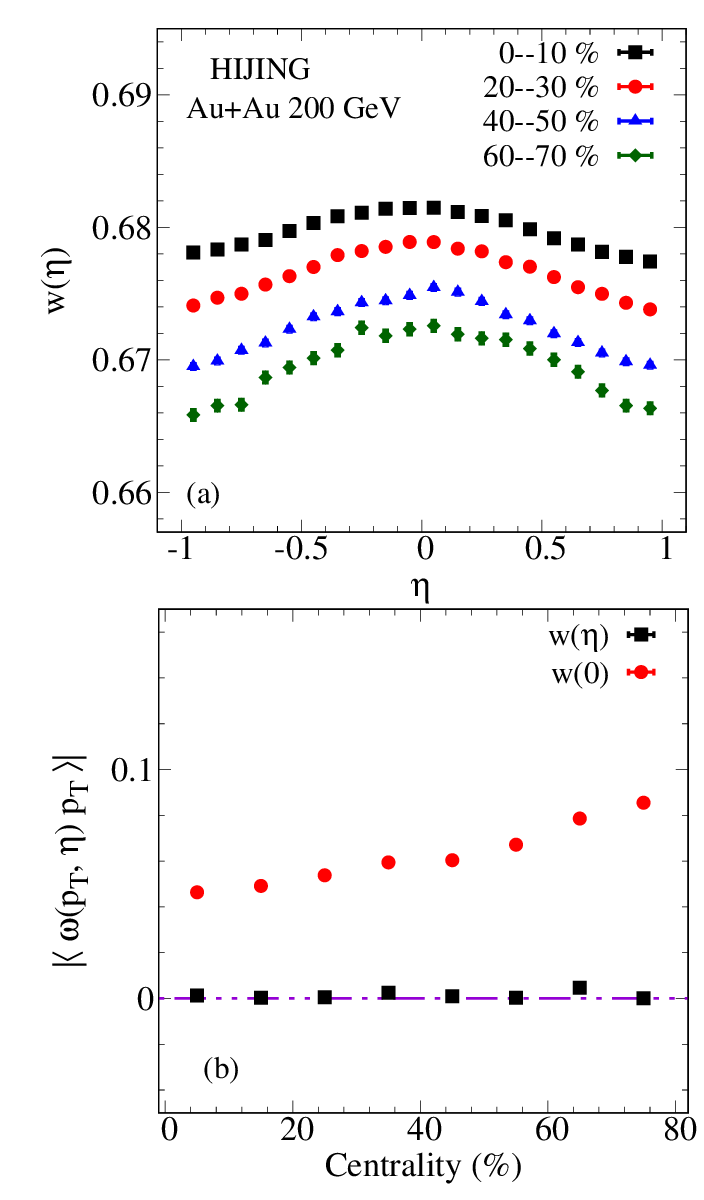}
\caption{
(a) The $\eta$ dependence of the weight $w(\eta)$ for several centrality intervals in HIJING Au+Au collisions at $\sqrt{s_{NN}}=200$ GeV. 
(b) Centrality dependence of $|\langle \omega(\eta, p_T)p_T\rangle|$ obtained using the $\eta$-dependent weight and a fixed midrapidity weight $w(0)$. The $\eta$-dependent procedure strongly suppresses the residual momentum-conservation contribution.
}
\label{fig:fig0}
\end{figure}
%-----------------------------------------------
%-----------------------------------------------
\begin{figure}[htb]
\centering
\includegraphics[width=0.9\linewidth]{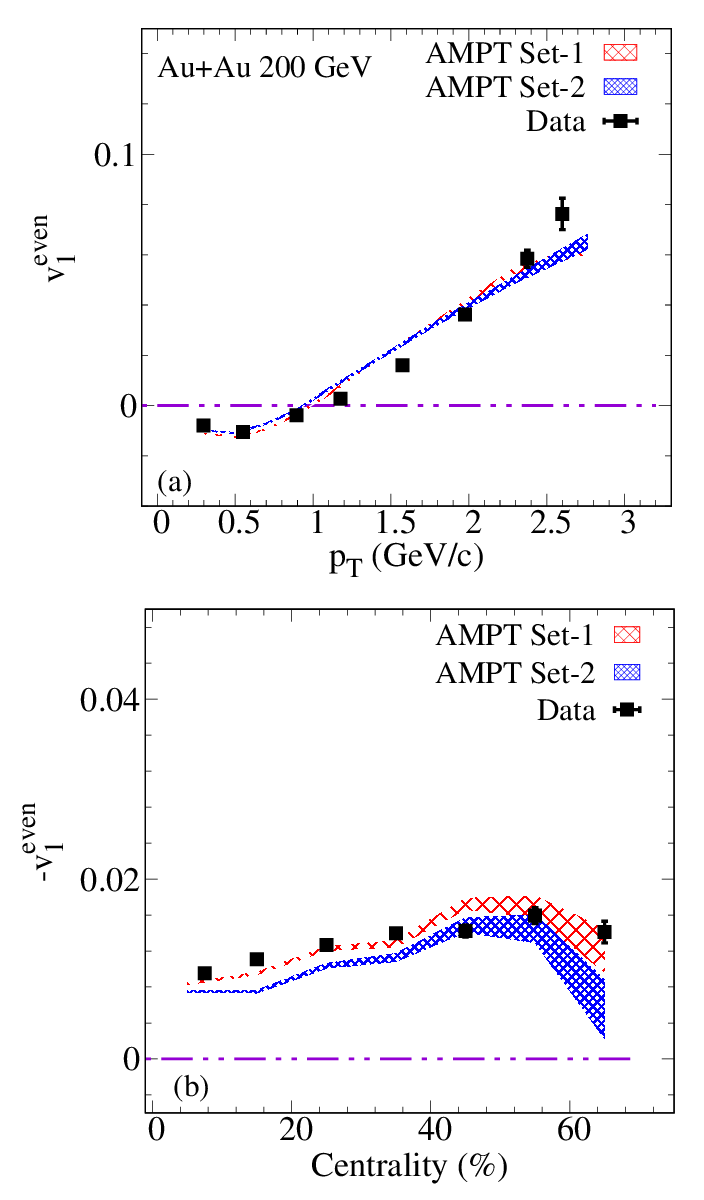}
\caption{
Comparison of AMPT Set-1 and Set-2 with available experimental data 
for Au+Au collisions at $\sqrt{s_{\rm NN}}=200$~GeV~\cite{STAR:2018gji}. 
(a) Transverse-momentum dependence of $v_1^{\rm even}$. 
(b) Centrality dependence of the integrated $-v_1^{\rm even}$ 
averaged over $0.4<p_T<0.7$~GeV/$c$.
}
\label{fig:v1comp}
\end{figure}
%-----------------------------------------------

%-----------------------------------------------
Figure~\ref{fig:fig0} illustrates the importance of retaining the $\eta$ dependence of the weight. The comparison with a fixed midrapidity weight, $w(0)$, is not intended as an independent proof of collective-flow extraction, since the orthogonality condition is largely built into the definition of $\omega$. Rather, it demonstrates that a single midrapidity or $\eta$-independent weight does not satisfy the same orthogonality condition away from midrapidity and therefore leaves a sizable residual momentum-conservation contribution. The more stringent validation of the procedure is provided by applying the full analysis chain to HIJING events, for which no collective final-state flow is expected.
%-----------------------------------------------

%-----------------------------------------------
The corrected first-harmonic vectors are used in the extraction of $v_1^{\rm even}$ and in all subsequent multi-particle correlations involving the first harmonic. This procedure is particularly important for the present study because the observables involving $v_1$ are designed to probe the genuine dipolar response and its correlation with elliptic and triangular flow, rather than correlations generated by global recoil.
%-----------------------------------------------

%-----------------------------------------------
To validate the weighting and suppression procedure, the AMPT results for $v_1^{\rm even}$ are compared with available experimental data. Figure~\ref{fig:v1comp} shows the comparison as a function of transverse momentum and centrality. The $p_T$ dependence exhibits the expected sign-changing behavior of rapidity-even dipolar flow, with negative values at low $p_T$ and positive values at higher $p_T$. The centrality dependence of the $p_T$-integrated signal is also reasonably reproduced by the AMPT calculation. These comparisons establish the model's baseline performance before the analysis is extended to multi-particle correlations.

%-----------------------------------------------
\subsubsection{Multi-particle correlations}
\label{subsec:correlators}
%-----------------------------------------------

The multi-particle correlation method is employed to quantify flow magnitudes, flow fluctuations, mixed-harmonic correlations, and event-plane correlations involving the rapidity-even dipolar harmonic~\cite{Bilandzic:2010jr, Jia:2017hbm, Huo:2017nms, Zhang:2018lls, Magdy:2020bhd, Alqahtani:2025vda}. The generic $k$-particle azimuthal correlator is defined as:
\begin{align}
C_{n_1,\ldots,n_k}|_{p_1, \ldots,p_k} &= 
\left\langle\left\langle
\cos\left(
n_1\phi_1+\cdots+n_k\phi_k
\right)
\right\rangle\right\rangle ,
\label{eq:general_corr}
\end{align}
where the inner average is taken over particles within an event, and the outer average is taken over events. The subscript notation $p_k$ is an index, where $0$ denotes an 
integrated particle of reference, while a nonzero index denotes the 
position of the differential particle of interest. For example, 
$C_{1-1}|_{10}$ denotes a two-particle correlator in which the first harmonic is differential in $p_T$ and the second harmonic is integrated over the reference-particle range. Rotational invariance requires $\sum_{i=1}^{k} n_i = 0$.

In the collective-flow picture, Eq.~\eqref{eq:general_corr} can be expressed as
\begin{align}
C_{n_1,n_2,\ldots,n_k}|_{p_1, \ldots,p_k}
&=
\left\langle
v_{|n_1|}\cdots v_{|n_k|}
\right.
\nonumber\\
&\qquad\left.
\cos\!\left(
n_1\Psi_{|n_1|}
+\cdots+
n_k\Psi_{|n_k|}
\right)
\right\rangle .
\label{eq:general_corr_flow}
\end{align}
Thus, depending on the harmonic combination, the same formal object can probe either flow magnitudes, flow fluctuations, or correlations among event-plane angles.
%---------------------------------------
%---------------------------------------
%---------------------------------------
\begin{table}[th]
\centering
\caption{Summary of the observables used in this work.}
\label{tab:observables}
\renewcommand{\arraystretch}{1.35}
\begin{tabular}{lc}
\hline
\hline
Observable & Definition \\
\hline
$\left\langle V_1V_{-1}\right\rangle$
&
$C_{1,-1}$ \\
\\

$\left\langle V_1V_1V_{-1}V_{-1}\right\rangle$
&
$C_{1,1,-1,-1}$ \\
\\

$\left\langle V_1V_2V_{-1}V_{-2}\right\rangle$
&
$C_{1,2,-1,-2}$ \\
\\

$\left\langle V_1V_3V_{-1}V_{-3}\right\rangle$
&
$C_{1,3,-1,-3}$ \\
\\

$\left\langle V_1V_1V_{-2}\right\rangle$ 
&
$C_{1,1,-2}$ \\
\\

$\left\langle V_1V_2V_{-3}\right\rangle$ 
&
$C_{1,2,-3}$ \\
\\

$\left\langle V_1V_1V_1V_{-3}\right\rangle$ 
&
$C_{1,1,1,-3}$ \\
\\

\(C_{1}\)\(\{4\}\)
&
$\left\langle V_1V_1V_{-1}V_{-1}\right\rangle
-
2\left\langle V_1V_{-1}\right\rangle^2$ \\
\\

$\beta_{1,2}$
&
$\dfrac{
\left\langle V_1V_2V_{-1}V_{-2}\right\rangle
}{
\left\langle V_1V_{-1}\right\rangle
\left\langle V_2V_{-2}\right\rangle
}
-1$ \\
\\

$\beta_{1,3}$
&
$\dfrac{
\left\langle V_1V_3V_{-1}V_{-3}\right\rangle
}{
\left\langle V_1V_{-1}\right\rangle
\left\langle V_3V_{-3}\right\rangle
}
-1$ \\
\\

$\rho_{1,2}$
&
$\dfrac{
\left\langle V_1V_1V_{-2}\right\rangle
}{
\sqrt{
\left|
\left\langle V_1V_1V_{-1}V_{-1}\right\rangle
\left\langle V_2V_{-2}\right\rangle
\right|}
}$ \\
\\

$\rho_{1,2,3}$
&
$\dfrac{
\left\langle V_1V_2V_{-3}\right\rangle
}{
\sqrt{
\left|
\left\langle V_1V_2V_{-1}V_{-2}\right\rangle
\left\langle V_3V_{-3}\right\rangle
\right|}
}$ \\
\\
\hline
\hline
\end{tabular}
\end{table}
Additional details of the correlator formalism and its $Q$-vector implementation can be found in Refs.~\cite{Magdy:2024ooh, Magdy:2024eci}.

For the two-particle correlations, a two-subevent method with $\Delta\eta>0.7$ is used to reduce short-range non-flow. The higher-order mixed-harmonic and asymmetric correlations are evaluated 
with the standard one-subevent method. This choice is motivated by the small magnitude of the first-harmonic signal after GMC suppression and the high statistical cost of imposing additional subevent requirements on multi-particle observables. Residual non-flow effects are therefore assessed using the same analysis chain applied to HIJING events. A dedicated multi-subevent study of the higher-order correlators would require substantially larger event samples and is left for future work.

%--------------------------------------------------------------------
%--------------------------------------------------------------------
%--------------------------------------------------------------------
%--------------------------------------------------------------------
\section{Results and discussion}\label{sec:4}
%--------------------------------------------------------------------
This section presents the centrality- and $p_{T}$-dependent results for rapidity-even dipolar flow and its multi-particle correlations in Au+Au collisions at $\sqrt{s_{NN}}=200$ GeV. The discussion is organized to isolate three distinct physics ingredients: (i) non-flow artifacts associated primarily with global momentum conservation (GMC), (ii) the collective response of the medium to the dipolar eccentricity $\varepsilon_1$, and (iii) the correlations of the dipole mode with elliptic and triangular flow. HIJING serves as a noncollective reference, while the two AMPT calculations are used to assess sensitivity to the partonic transport setting. Changing the partonic scattering strength can modify the mean harmonic response, the event-by-event harmonic distributions, magnitude correlations, and angular correlations. Consequently, the Set-1/Set-2 differences are interpreted as transport-setting dependence and not as a direct measure of viscous damping alone. A comparison of event activity between the AMPT and HIJING models is presented in Appendix~\ref{app:Comparison}.

%--------------------------------------------------------------------
%--------------------------------------------------------------------
\begin{figure}[th]
\centering
\includegraphics[width=1.0\linewidth]{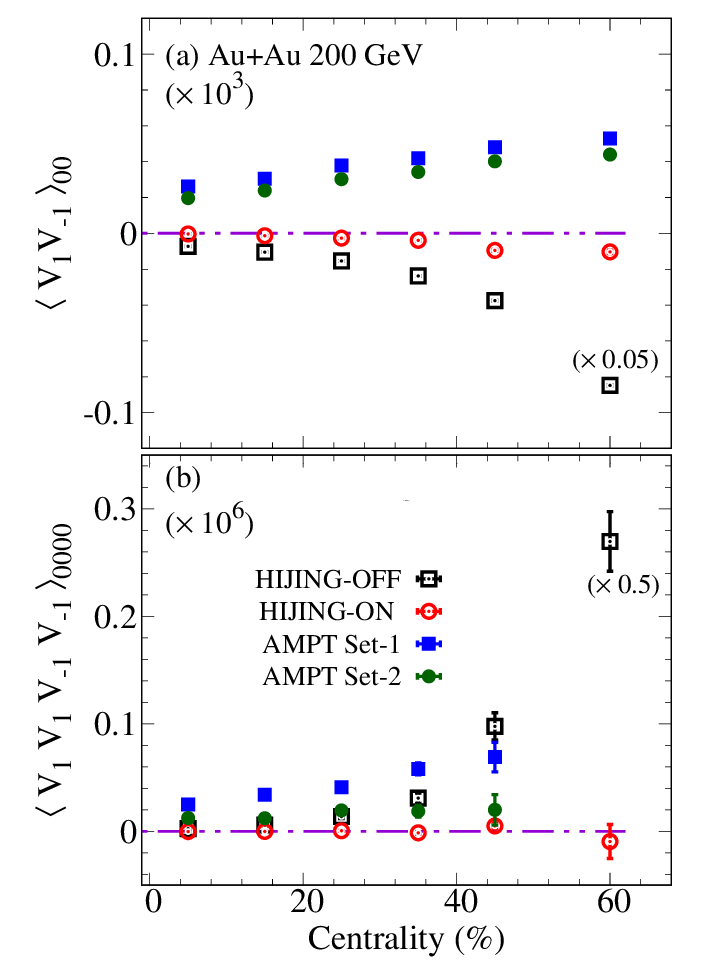}
\caption{
Centrality dependence of the integrated two-particle correlation $\langle V_1V_{-1}\rangle_{00}$ panel (a) and the four-particle correlation $\langle V_1V_1V_{-1}V_{-1}\rangle_{0000}$ panel (b). The HIJING results are shown before GMC suppression (HIJING-OFF) and after GMC suppression (HIJING-ON). The AMPT results are shown for two transport parameter sets as given in Tab~\ref{tab:1}. The numerical scale factors shown in the panels are applied only for visual clarity.
\label{fig:v1_integrated}
}
\end{figure}
%--------------------------------------------------------------------
\begin{figure}[h]
\centering
\includegraphics[width=1.0\linewidth]{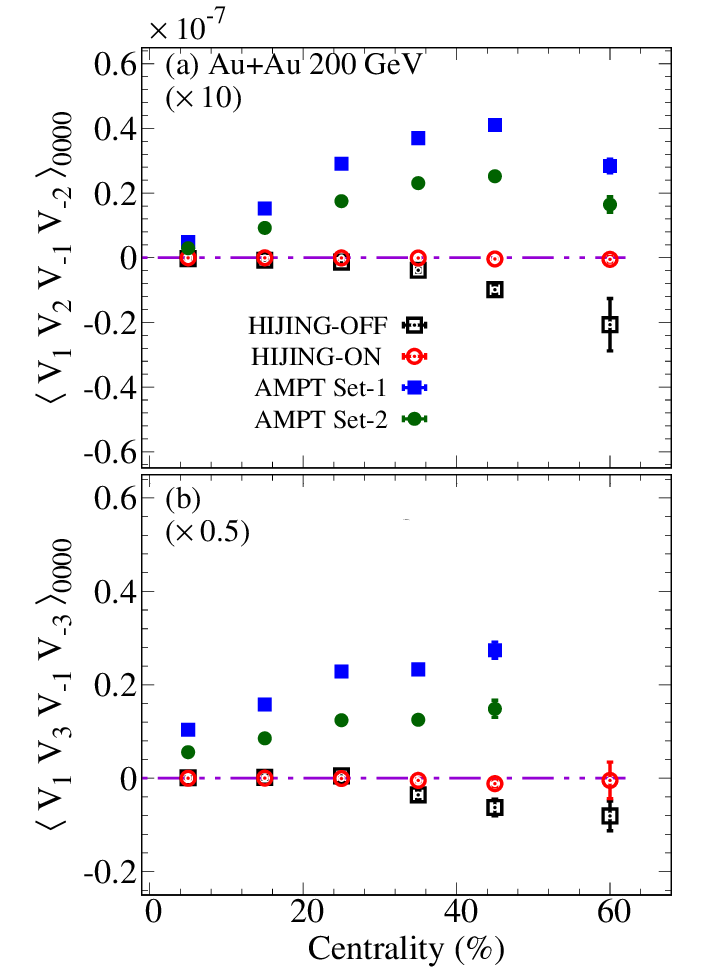}
\caption{
Same as in Fig.~\ref{fig:v1_integrated}, but for $\langle V_1V_2V_{-1}V_{-2}\rangle_{0000}$ (a) and $\langle V_1V_3V_{-1}V_{-3}\rangle_{0000}$ (b).
\label{fig:sc_integrated}
}
\end{figure}
%--------------------------------------------------------------------
\begin{figure}[h]
\centering
\includegraphics[width=1.0\linewidth]{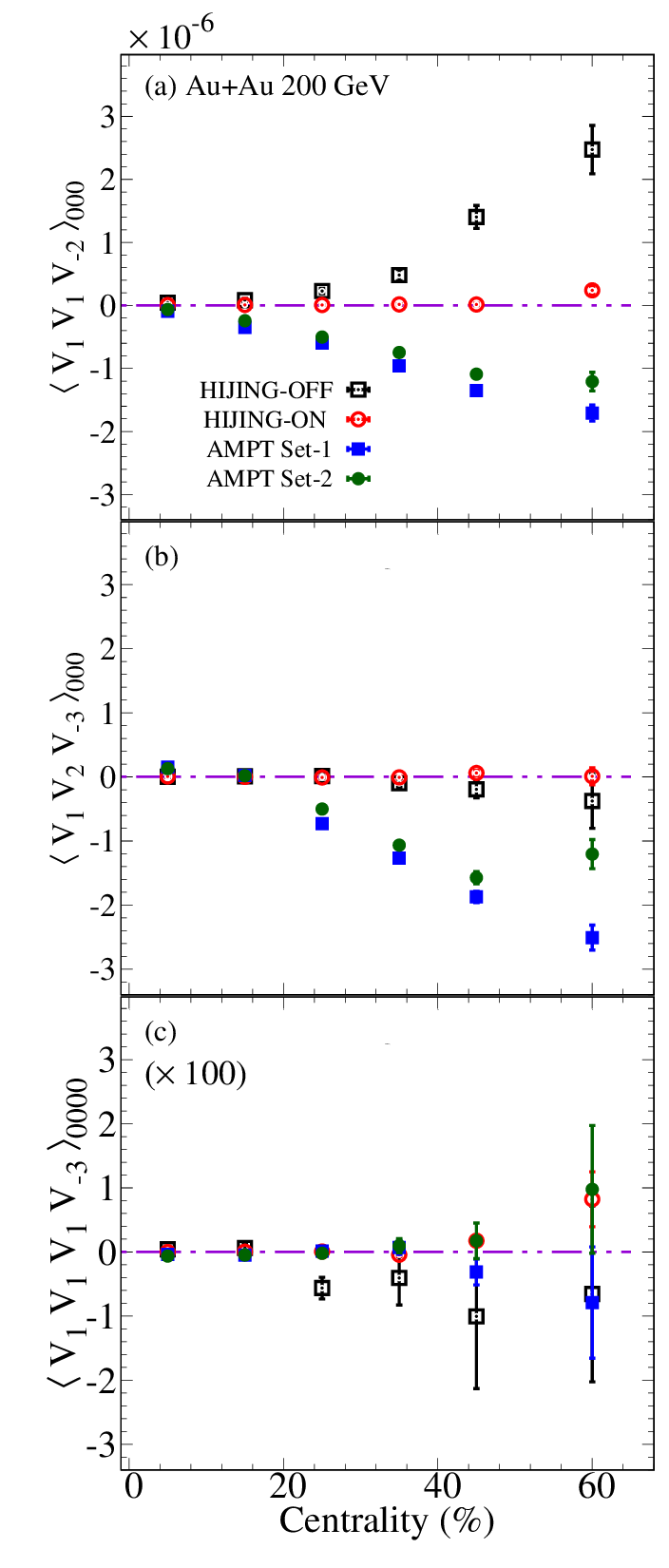}
\caption{
Same as in Fig.~\ref{fig:v1_integrated}, but for $\langle V_1V_1V_{-2}\rangle_{000}$ (a), $\langle V_1V_2V_{-3}\rangle_{000}$ (b), and $\langle V_1V_1V_1V_{-3}\rangle_{0000}$ (c).
\label{fig:asc_integrated}
}
\end{figure}
%--------------------------------------------------------------------

%--------------------------------------------------------------------
Figure~\ref{fig:v1_integrated} shows the centrality dependence of the integrated two- and four-particle correlation associated with rapidity-even directed flow. These observables provide the first direct test of the GMC suppression. The GMC effect is clearly seen in the HIJING-OFF results, which exhibit a sizable centrality-dependent contribution, even though HIJING lacks a collective hydrodynamic response. The magnitude of this raw contribution grows toward peripheral collisions, consistent with the expected enhancement of momentum-conservation effects at lower multiplicity~\cite{STAR:2018gji}. After the GMC suppression is applied, the HIJING-ON results are strongly suppressed and remain close to zero over the presented centrality range. This behavior demonstrates that the $\eta$-dependent weighting procedure removes the leading GMC artifact from the first-harmonic moments. The small residual deviation from zero in $\langle V_1V_{-1}\rangle_{00}$ indicates that two-particle observables can still retain contributions from residual long-range non-flow correlations, such as dijet-like recoil that are not fully eliminated by the suppression weight. The corresponding four-particle correlation is more strongly suppressed, reflecting the well-known advantage of higher-order moments in reducing few-particle non-flow effects.

The AMPT results differ qualitatively from the HIJING-ON baseline. Both the two- and four-particle first-harmonic correlations show nonzero signals with a systematic dependence on the AMPT partonic transport setting. This sensitivity indicates that the extracted correlations are not merely residual non-flow within the HIJING baseline, but are also affected by the change in the partonic transport setting within the AMPT model.
%--------------------------------------------------------------------
%--------------------------------------------------------------------

%--------------------------------------------------------------------
%--------------------------------------------------------------------
Figure~\ref{fig:sc_integrated} presents the integrated mixed-harmonic four-particle correlation observables, which probe correlations between the magnitude of rapidity-even dipolar flow and the magnitudes of higher-order flow harmonics. The comparison between HIJING-OFF and HIJING-ON again illustrates the importance of the GMC suppression weight. 
The AMPT calculations exhibit a pronounced dependence on $\sigma_{\rm parton}$. The larger $\sigma_{\rm parton}$ calculation yields larger mixed-harmonic correlations, whereas the lower $\sigma_{\rm parton}$ set yields systematically lower values. This ordering is expected if the mixed correlations are sensitive to the interplay between initial- and final-state effects.

%--------------------------------------------------------------------
%--------------------------------------------------------------------
%--------------------------------------------------------------------
Figure~\ref{fig:asc_integrated} shows the integrated asymmetric correlations involving $V_1$, $V_2$, and $V_3$. These observables are sensitive to correlations among the event-plane angles. They therefore provide complementary information to the magnitude correlations: two harmonics may have correlated magnitudes because they originate from a common fluctuation strength, while their event-plane orientations may remain weakly correlated if the corresponding spatial structures are independently oriented.

The observable $\langle V_1V_1V_{-2}\rangle_{000}$ in Fig.~\ref{fig:asc_integrated}(a) is proportional to an event-plane correlation of the form $\langle v_1^2v_2\cos(2\Psi_1-2\Psi_2)\rangle$. The three-plane observable $\langle V_1V_2V_{-3}\rangle_{000}$, shown in Fig.~\ref{fig:asc_integrated}(b), probes the rotationally invariant combination $\cos(\Psi_1+2\Psi_2-3\Psi_3)$. The AMPT calculations show a pronounced signal with a clear centrality dependence. The sign and magnitude indicate a robust angular correlation between $\Psi_1$ and $\Psi_2$ for (a) and between $\Psi_1$, $\Psi_2$, and $\Psi_3$. Both observables show moderate sensitivity to variations in $\sigma_{\rm parton}$.

The direct dipole--triangular angular correlation is tested by $\langle V_1V_1V_1V_{-3}\rangle_{0000}$ in Fig.~\ref{fig:asc_integrated}(c), which corresponds to $\langle v_1^3v_3\cos(3\Psi_1-3\Psi_3)\rangle$. This observable remains close to zero within uncertainties, indicating that AMPT does not generate a strong direct locking between the dipolar and triangular event-plane angles in this channel. This result does not exclude correlations among the dipolar, elliptic, and triangular planes. In particular, the coupled three-plane structure $\cos(\Psi_1+2\Psi_2-3\Psi_3)$ is probed separately by $\langle V_1V_2V_{-3}\rangle$ and by the normalized observable.
%--------------------------------------------------------------------
%--------------------------------------------------------------------
%--------------------------------------------------------------------

%--------------------------------------------------------------------
%--------------------------------------------------------------------
%--------------------------------------------------------------------
\begin{figure}[h]
\centering
\includegraphics[width=1.0\linewidth]{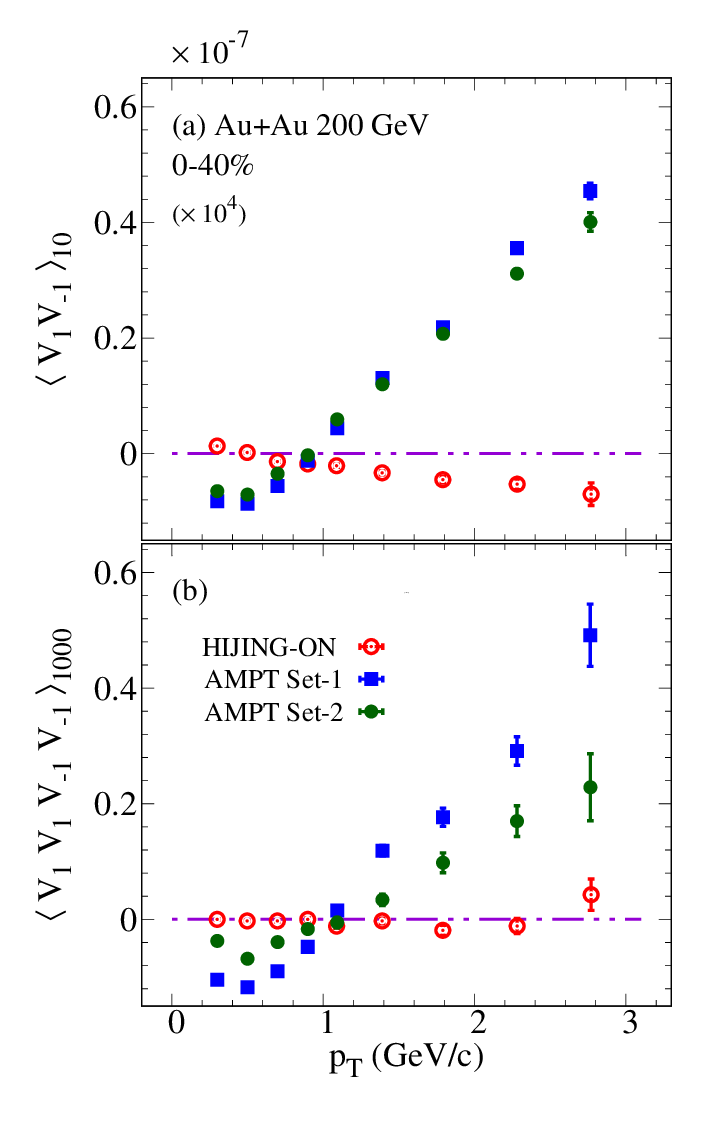}
\caption{
Transverse-momentum dependence of the two-particle correlation $\langle V_1V_{-1}\rangle_{10}$ (a) and the four-particle correlation $\langle V_1V_1V_{-1}V_{-1}\rangle_{1000}$ (b) for $0$--$40\%$ Au+Au collisions at $\sqrt{s_{NN}}=200$ GeV. The HIJING result is shown only after GMC suppression, while AMPT Set-1 and Set-2 correspond to different $\sigma_{\rm parton}$ values as given in Tab~\ref{tab:1}. Scale factors are indicated in the panels.
\label{fig:v1_diff}
}
\end{figure}
%--------------------------------------------------------------------
\begin{figure}[h]
\centering
\includegraphics[width=1.0\linewidth]{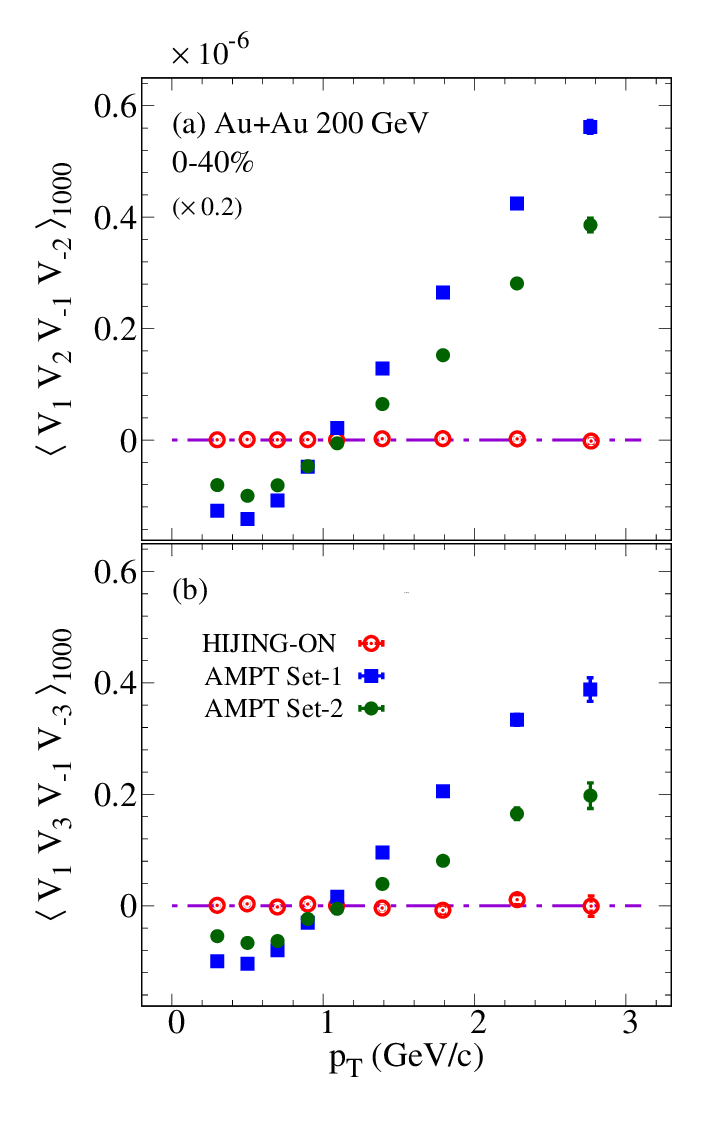}
\caption{
Same as in Fig.~\ref{fig:v1_diff}, but for $\langle V_1V_2V_{-1}V_{-2}\rangle_{1000}$ (a) and $\langle V_1V_3V_{-1}V_{-3}\rangle_{1000}$ (b).
\label{fig:diff_v1_mixed}
}
\end{figure}
%--------------------------------------------------------------------
\begin{figure}[h]
\centering
\includegraphics[width=1.0\linewidth]{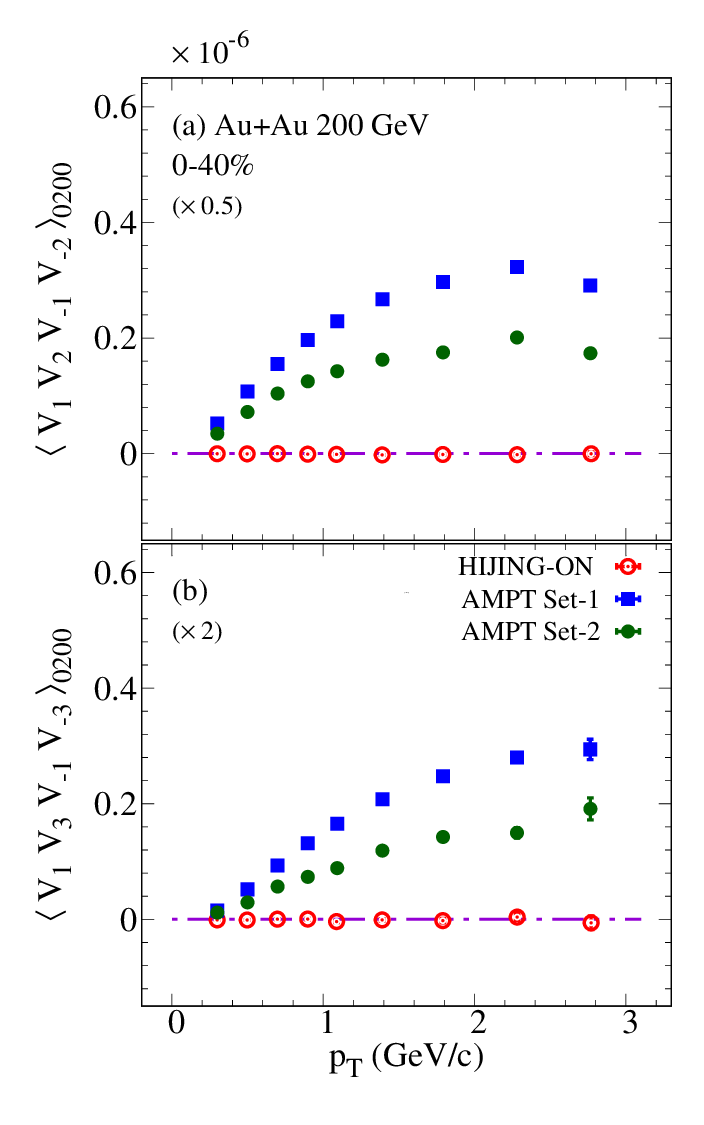}
\caption{
Same as in Fig.~\ref{fig:v1_diff}, but for $\langle V_1V_2V_{-1}V_{-2}\rangle_{0200}$ (a) and $\langle V_1V_3V_{-1}V_{-3}\rangle_{0200}$ (b).
\label{fig:diff_v23_mixed}
}
\end{figure}
%--------------------------------------------------------------------
\begin{figure}[h]
\centering
\includegraphics[width=1.0\linewidth]{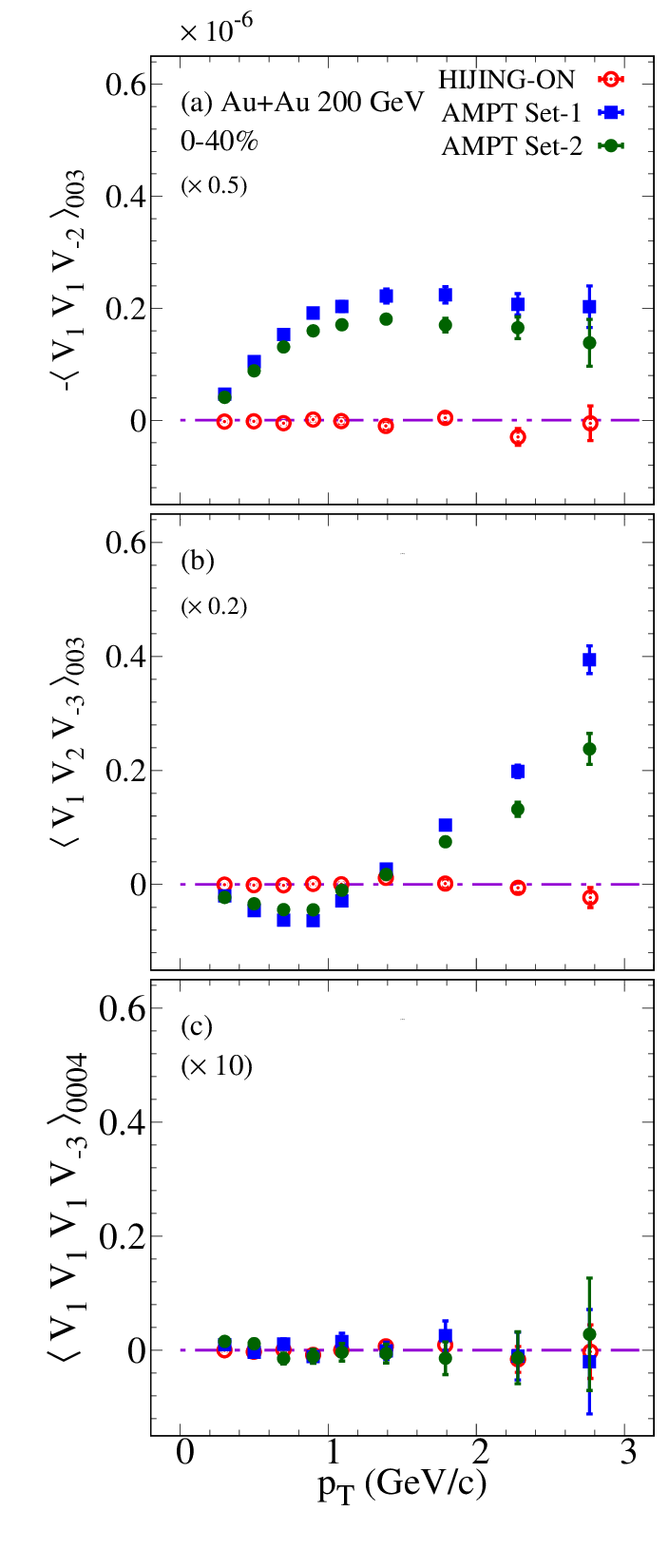}
\caption{
Same as in Fig.~\ref{fig:v1_diff}, but for $-\langle V_1V_1V_{-2}\rangle_{003}$ (a), $\langle V_1V_2V_{-3}\rangle_{003}$ (b) and $\langle V_1V_1V_1V_{-3}\rangle_{0004}$ (c).
\label{fig:diff_asc}
}
\end{figure}
%--------------------------------------------------------------------

%--------------------------------------------------------------------
%--------------------------------------------------------------------
Figure~\ref{fig:v1_diff} shows the $p_T$ dependence of the $V_{1}$ two- and four-particle correlation. For the $p_T$ dependence observables, only the GMC-suppressed HIJING baseline is shown. The differential two-particle correlation $\langle V_1V_{-1}\rangle_{10}$ exhibits the characteristic $p_T$ structure of rapidity-even dipolar flow. The AMPT results are negative at low $p_T$, cross zero at intermediate $p_T$, and become positive at higher $p_T$. This sign-changing behavior is a defining feature of $v_1^{\rm even}$ and follows from the constraint that the total transverse momentum generated by the dipolar flow field must vanish. Hydrodynamic calculations predict this behavior precisely: soft particles recoil opposite to the direction of the harder component so that the net transverse momentum remains zero~\cite{Teaney:2010vd, Retinskaya:2012ky}. The present AMPT calculations reproduce this qualitative structure, reinforcing the interpretation of the signal as a collective dipolar response rather than a conventional rapidity-odd directed-flow component.

The comparison between the two AMPT sets shows clear sensitivity to the partonic transport setting. The calculation with the larger $\sigma_{\rm parton}$ produces a stronger dipolar response, particularly at high $p_T$, whereas the smaller-$\sigma_{\rm parton}$ result is systematically reduced. This ordering reflects differences in the efficiency with which the initial dipolar asymmetry is converted into final-state momentum anisotropy. Because $v_{1}^{\rm even}(p_T)$ is constrained to change sign, the transport dependence affects not only its overall magnitude but also the relative balance between the low- and high-$p_T$ regions.
%--------------------------------------------------------------------
%--------------------------------------------------------------------

%--------------------------------------------------------------------
%--------------------------------------------------------------------
Figures~\ref{fig:diff_v1_mixed} and~\ref{fig:diff_v23_mixed} present complementary $p_T$-differential projections of the mixed-harmonic correlations. In Fig.~\ref{fig:diff_v1_mixed}, the particle of interest is assigned to the dipolar harmonic, whereas in Fig.~\ref{fig:diff_v23_mixed} the differential particle is assigned to the higher-order harmonic, $V_2$ or $V_3$. This comparison provides a direct way to identify which harmonic controls the observed $p_T$ dependence of the mixed correlation. In addition, the GMC-suppressed HIJING results remain close to zero in both figures, indicating that the observed $p_T$ dependence is not reproduced by the recoil and few-particle non-flow correlations included in the HIJING baseline.

When the differential harmonic is associated with $V_1$, both the $(1,2)$ and $(1,3)$ correlations exhibit the characteristic sign-changing structure of rapidity-even dipolar flow: they are negative at low $p_T$, cross zero near the expected sign-change region of $v_1^{\rm even}(p_T)$, and become positive at higher $p_T$. This indicates that the $p_T$ dependence in Fig.~\ref{fig:diff_v1_mixed} is controlled primarily by the differential $v_1^{\rm even}$ component. By contrast, when the differential particle is associated with $V_2$ or $V_3$, as shown in Fig.~\ref{fig:diff_v23_mixed}, the mixed correlations follow the more conventional hydrodynamic $p_T$ dependence of elliptic and triangular flow. The $(1,2)$ correlation increases monotonically with $p_T$ before reaching a broad saturation, reflecting the growth of elliptic anisotropy in the soft-to-intermediate-$p_T$ region, while the $(1,3)$ correlation shows a similar triangular-flow-like rise. The absence of a sign change in these projections demonstrates that the sign-changing pattern seen in Fig.~\ref{fig:diff_v1_mixed} is not an intrinsic property of the mixed correlator itself, but rather originates from the differential rapidity-even dipolar harmonic.

Both Figs.~\ref{fig:diff_v1_mixed} and~\ref{fig:diff_v23_mixed} exhibit a clear ordering between the two AMPT transport settings, with the larger partonic cross section producing mixed-harmonic correlations of greater magnitude. This ordering demonstrates sensitivity to the strength of the final-state interactions. Because these observables involve products of several harmonic amplitudes, their dependence on the transport setting can be stronger than that of single-harmonic measurements. 
%--------------------------------------------------------------------
%--------------------------------------------------------------------
%--------------------------------------------------------------------

%--------------------------------------------------------------------
%--------------------------------------------------------------------
%--------------------------------------------------------------------
%--------------------------------------------------------------------
%--------------------------------------------------------------------
Figure~\ref{fig:diff_asc} presents differential asymmetric correlations that probe the $p_T$ dependence of event-plane couplings involving the dipole mode. For the differential asymmetric correlations, the closing-harmonic convention is used. A three-particle correlator $C_{n,m,-(n+m)}=\langle V_{n} V_{m}  V_{-(n+m)} \rangle$ is evaluated with the first two harmonic entries integrated and the differential particle assigned to the closing harmonic. Assigning the differential particle to one of the input harmonics would define a distinct differential projection, which is beyond the scope of this work. The HIJING-ON results are flat and consistent with zero in all panels, demonstrating that the GMC-suppressed non-collective baseline does not generate sizable differential angular correlations.

The observable $-\langle V_1V_1V_{-2}\rangle_{003}$ in Fig.~\ref{fig:diff_asc}(a) probes the differential coupling between the dipolar and elliptic planes when the differential harmonic is associated with the elliptic harmonic. The correlation increases with $p_T$ and then approaches saturation, consistent with the expected growth of elliptic anisotropy in the soft sector.  Panel (b) shows $\langle V_1V_2V_{-3}\rangle_{003}$, which is sensitive to the coupled angular structure among $\Psi_1$, $\Psi_2$, and $\Psi_3$ with the differential dependence carried by the triangular harmonic. This observable shows a particularly interesting transition from small or slightly negative values at low $p_T$ to positive values at higher $p_T$. The shape resembles the characteristic dipolar response even though the differential harmonic is associated with $V_3$~\cite{Teaney:2010vd}. This indicates that the correlation is not solely due to the monotonic growth of $v_3(p_T)$; rather, it reflects the coupled response of the dipolar, elliptic, and triangular modes. The dependence on $\sigma_{\rm parton}$ in panels (a) and (b) is weaker than in the magnitude-based correlations but remains visible. 

The four-particle asymmetric correlation $\langle V_1V_1V_1V_{-3}\rangle_{0004}$ in Fig.~\ref{fig:diff_asc}(c) remains close to zero within uncertainties. This result is consistent with the integrated observation that the direct $\Psi_1$--$\Psi_3$ correlation is weak. Although the magnitudes of $v_1$ and $v_3$ can be correlated through common hotspot fluctuations, their symmetry-plane orientations are largely decorrelated. The weak signal in this channel, therefore, supports a picture in which the dipole and triangular modes share fluctuation strength but do not exhibit strong direct angular locking.
%--------------------------------------------------------------------

%--------------------------------------------------------------------
\begin{figure}[t]
\centering
\includegraphics[width=1.0\linewidth]{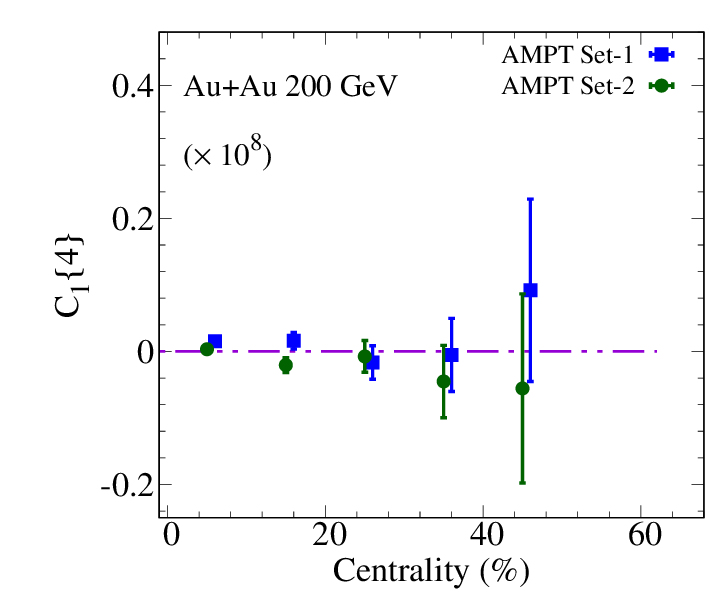}
\caption{
Centrality dependence of the first-harmonic cumulant, \(C_{1}\)\(\{4\}\), observable. The results are shown for two AMPT transport parameter sets.
\label{fig:Cum}
}
\end{figure}
%--------------------------------------------------------------------
%--------------------------------------------------------------------
Figure~\ref{fig:Cum} shows the centrality dependence of the first-harmonic four-particle cumulant, $C_{1}\{4\}$. The cumulant remains consistent with zero within the present uncertainties over the measured centrality range. This indicates that the four-particle contribution is small compared with the underlying second- and fourth-order first-harmonic moments and cannot be resolved with the available statistical precision. The result is qualitatively reminiscent of the small four-particle cumulant observed for the fluctuation-dominated triangular harmonic~\cite{STAR:2013qio}.
However, the near-zero value of $C_{1}\{4\}$ alone does not determine
the detailed form of the event-by-event $v_{1}^{\rm even}$
distribution, because it may also result from a cancellation between the two terms entering the cumulant.
%--------------------------------------------------------------------

%--------------------------------------------------------------------
%--------------------------------------------------------------------
%--------------------------------------------------------------------
Previous experimental studies have shown that normalized symmetric and asymmetric correlations exhibit reduced sensitivity to the overall magnitudes of the final-state flow response and can therefore provide a cleaner measure of the relative correlations among the magnitudes and orientations of different flow harmonics~\cite{Magdy:2018itt, Magdy:2020bij, STAR:2022vkx, STAR:2022gki}. Motivated by these observations, the corresponding normalized observables in AMPT are examined, and their sensitivity to final-state transport is tested by varying the partonic scattering strength. The comparison between the two AMPT transport settings determines the extent to which normalization reduces the dependence on the absolute final-state response while preserving sensitivity to the underlying harmonic correlations and fluctuations. The normalized final-state observables are therefore presented, together with their initial-state counterparts, as functions of centrality. This comparison tests whether the initial correlation structure is preserved and how it is modified by the subsequent medium evolution.
%--------------------------------------------------------------------

%--------------------------------------------------------------------
For $p_T$-differential measurements, however, the normalized ratios involve small and strongly fluctuating numerators and denominators, and the available event samples do not provide sufficient statistical precision for a stable extraction. The normalized analysis is therefore restricted to centrality-dependent integrated observables, while the $p_T$-differential correlations are presented in non-normalized form.
%--------------------------------------------------------------------

%--------------------------------------------------------------------
\begin{figure}[t]
\centering
\includegraphics[width=1.0\linewidth]{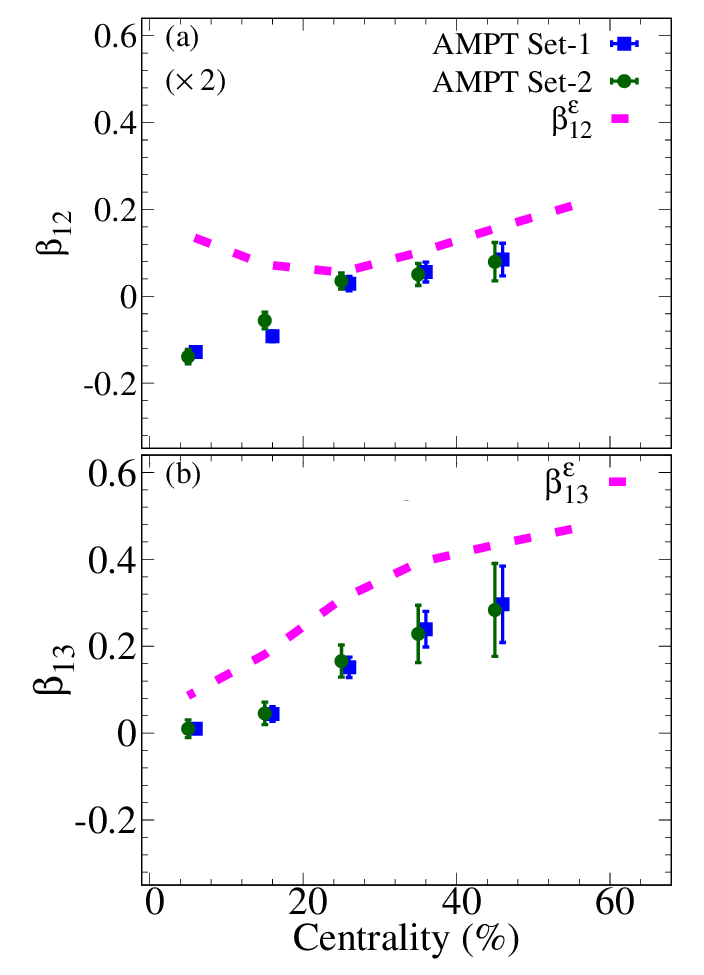}
\caption{
Centrality dependence of the normalized mixed-harmonic correlations $\beta_{12}$ (a) and $\beta_{13}$ (b). The results are shown for two AMPT transport parameter sets.  The dashed curve shows the initial-state eccentricity calculations.
\label{fig:beta}
}
\end{figure}
%--------------------------------------------------------------------

%--------------------------------------------------------------------
%--------------------------------------------------------------------
Figure~\ref{fig:beta} compares the normalized final-state mixed-harmonic correlations $\beta_{12}$ and $\beta_{13}$ with their corresponding initial-state eccentricity correlations. The final-state results exhibit weak sensitivity to the AMPT partonic transport setting. This behavior indicates that normalization removes a large part of the dependence on the absolute harmonic magnitudes and their average response coefficients.
The comparison with the eccentricity correlations then tests how the initial correlation structure is preserved or modified during the medium evolution.

The final-state $\beta_{12}$ is negative in central collisions, crosses zero at intermediate centrality, and becomes positive toward peripheral collisions. The negative values correspond to an
anti-correlation between $(v_{1}^{even})^2$ and $(v_{2})^2$, whereas the positive values indicate a positive correlation between their magnitudes. In contrast, the initial-state
$\beta^{\varepsilon}_{12}$ remains positive over the studied centrality range and generally increases toward peripheral collisions. Thus, the increasing peripheral trend is already present
in the initial geometry, but the negative central values and the sign change of the final-state observable are not inherited directly from the initial eccentricity correlation. Instead, they indicate that the initial dipolar--elliptic magnitude correlation is modified by the medium response, particularly in central collisions.

The $\beta_{13}$ correlation is positive and increases toward peripheral collisions in both the initial and final states. The similar centrality dependence indicates that the positive correlation
between the dipolar and triangular fluctuation strengths originates primarily from the initial geometry and is preserved through the subsequent transport evolution. The smaller magnitude of the
final-state $\beta_{13}$ relative to $\beta^{\varepsilon}_{13}$ shows, however, that the strength of this correlation is modified by the medium response.
%--------------------------------------------------------------------
%--------------------------------------------------------------------

%--------------------------------------------------------------------
\begin{figure}[t]
\centering
\includegraphics[width=1.0\linewidth]{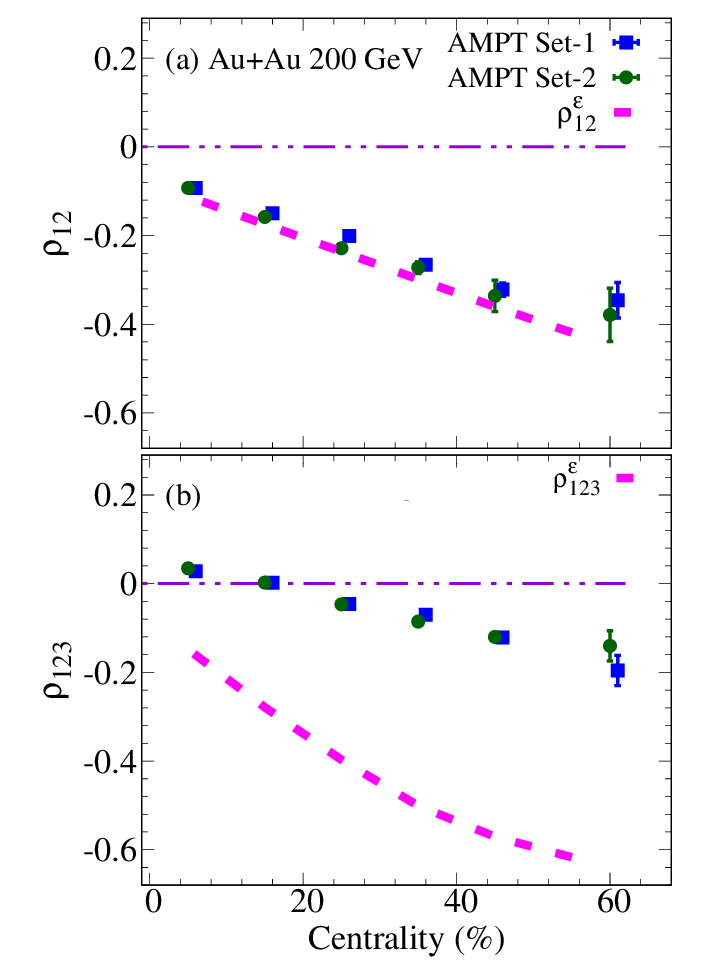}
\caption{
Centrality dependence of normalized event-plane correlations in Au+Au collisions at $\sqrt{s_{NN}}=200$ GeV. Panel (a) shows $\rho_{12}$, which quantifies the correlation between $\Psi_1$ and $\Psi_2$. Panel (b) shows $\rho_{123}$, which quantifies the event-plane correlation among $\Psi_1$, $\Psi_2$, and $\Psi_3$. Results are shown for two AMPT transport parameter sets. The dashed curve shows the initial-state eccentricity calculations.
\label{fig:rho}
}
\end{figure}
%--------------------------------------------------------------------
%--------------------------------------------------------------------

%--------------------------------------------------------------------
Figure~\ref{fig:rho} presents the normalized angular correlations $\rho_{12}$ and $\rho_{123}$ together with their corresponding initial-state eccentricity correlations. Normalization by appropriate harmonic-magnitude moments reduces dependence on absolute flow amplitudes, allowing a more direct comparison of angular-correlation structures between the initial and final states.

The $\rho_{12}$ observable is negative over the full centrality range and becomes increasingly negative toward peripheral collisions. The corresponding initial-state correlation exhibits a similar sign, magnitude, and centrality dependence. This close agreement indicates that the anti-correlation between the dipolar and elliptic orientations is predominantly established by the initial collision geometry and is largely preserved during the final-state evolution. Its increasing magnitude toward peripheral collisions is consistent with the growing influence of the almond-shaped overlap geometry, which establishes a preferred elliptic direction and constrains the orientation of the dipolar fluctuation.

The three-harmonic correlation $\rho_{123}$ is slightly positive in the most central events and becomes negative with increasing centrality. By contrast, the corresponding initial-state correlation is already negative in central collisions and becomes progressively more negative toward the peripheral region. Therefore, the negative midcentral-to-peripheral trend is present in the initial geometry. In addition, the final-state correlation has an absolute magnitude substantially smaller than its initial-state counterpart. These differences show that the medium modifies the strength of the three-harmonic angular correlation, even though much of the underlying centrality dependence remains visible in the final state. The weak separation between the two AMPT transport settings indicates that the normalized observable is only weakly sensitive to variations in the partonic scattering strength.
%--------------------------------------------------------------------
%--------------------------------------------------------------------

%--------------------------------------------------------------------
%--------------------------------------------------------------------
The comparison between the initial- and final-state normalized observables clarifies the role of normalization in reducing the dependence on the final-state response. The weak separation between the two AMPT transport settings demonstrates that the normalized $\beta$ and $\rho$ observables are considerably less sensitive to the absolute response magnitudes than the corresponding non-normalized correlations. This reduced dependence on transport settings does not imply that the normalized final-state observables are identical to their initial-state counterparts. Rather, they reflect the initial eccentricity correlations after those correlations have been transformed by the medium response.

The level of preservation depends on the observable. The centrality dependence of $\rho_{12}$ is transferred from the initial to the final state with little modification, while $\beta_{13}$ retains the initial trend with a reduced correlation strength. In contrast, the central behavior of $\beta_{12}$ and $\rho_{123}$ is qualitatively modified during the evolution. The normalized observables, therefore, provide sensitivity both to the underlying initial-state correlation
structure and to medium-induced modifications of that structure, while suppressing much of the leading dependence on the individual flow magnitudes.

Experimental measurements of GMC-suppressed $v_{1}^{\rm even}$ correlations, including the non-normalized symmetric and asymmetric correlations and their normalized $\beta$ and $\rho$ counterparts, would consequently provide complementary constraints on the magnitude and correlation structure of the dipolar response. Direct comparisons with model calculations can test whether a given description simultaneously reproduces the initial eccentricity correlations, the final-state harmonic magnitudes, and the medium-driven transformation of the correlations between them.
%--------------------------------------------------------------------
%--------------------------------------------------------------------

%--------------------------------------------------------------------
Future work should extend this analysis to additional collision systems, beam energies, and initial-condition models. Since the dipolar eccentricity is strongly weighted toward the nuclear surface, $v_1^{\rm even}$ and its correlations are expected to be sensitive to surface diffuseness, neutron skins, and clustered nuclear configurations, making isobaric systems, light-ion collisions such as ${}^{16}{\rm O}+{}^{16}{\rm O}$, and small systems especially valuable.
%--------------------------------------------------------------------

%--------------------------------------------------------------------
\section{Summary and Conclusions} \label{sec:5}
\label{sec:summary}

A systematic study of rapidity-even dipolar flow, $v_1^{\rm even}$, and its multi-particle correlations in Au+Au collisions at $\sqrt{s_{NN}}=200$~GeV is presented using AMPT and HIJING. An $\eta$-dependent weight is applied to suppress the leading global-momentum-conservation contribution, after which the HIJING results are strongly reduced and remain close to zero for most integrated and differential observables. The AMPT calculations reproduce the characteristic sign change of $v_1^{\rm even}(p_T)$, with negative values at low $p_T$ and positive values at higher $p_T$, supporting its interpretation as a collective response to the initial dipolar asymmetry. The non-normalized first-harmonic and mixed-harmonic correlations show clear sensitivity to the AMPT partonic transport setting, reflecting changes in the harmonic magnitudes, their event-by-event fluctuations, and their mutual correlations. By contrast, the normalized observables exhibit much weaker dependence on $\sigma_{\rm parton}$, indicating that normalization removes a substantial part of the leading dependence on the absolute final-state response. Direct comparison with the corresponding initial-state calculations further shows that these normalized observables retain the underlying initial-state correlation structure after modification by the medium. Together, these results demonstrate that GMC-suppressed $v_1^{\rm even}$ correlations provide complementary sensitivity to the initial dipolar geometry and to its transformation during final-state evolution.
%--------------------------------------------------------------------

%--------------------------------------------------------------------
\section*{Acknowledgments}
%--------------------------------------------------------------------
%, and Jean-Yves Ollitrault
%The author thanks Jiangyong Jia, Jean-Yves Ollitrault, Mubarak Alqahtani, and Chunjian Zhang for the valuable discussions and for pointing out essential references. The author acknowledges using ChatGPT (OpenAI) for language revision and text improvement.
%This work was performed by using computational resources offered by the High-Performance Computing Center at Texas Southern University.

The author thanks Jiangyong Jia, Jean-Yves Ollitrault, Mubarak Alqahtani, and Chunjian Zhang for valuable discussions and pointing out essential references, and acknowledges the use of ChatGPT (OpenAI) for language revision and text improvement. This work was performed using computational resources offered by the High-Performance Computing Center at Texas Southern University. %The codes used to generate the simulated data from the AMPT and HIJING models are publicly available on GitHub~\cite{PC_NPC_In_AMPT_2026, PC_NPC_In_HIJING_2026}.

%--------------------------------------------------------------------
%--------------------------------------------------------------------
%--------------------------------------------------------------------
%\clearpage

\appendix
%--------------------------------------------------------------------
\input{appendix_1.tex}

%--------------------------------------------------------------------
%\clearpage
%\begin{widetext}
%--------------------------------------------------------------------
%--------------------------------------------------------------------
%--------------------------------------------------------------------
%\input{appendix_eq.tex}
%--------------------------------------------------------------------
%--------------------------------------------------------------------
%\end{widetext}
%--------------------------------------------------------------------
%--------------------------------------------------------------------
%--------------------------------------------------------------------

%--------------------------------------------------------------------
%--------------------------------------------------------------------
%--------------------------------------------------------------------
%\bibliographystyle{aipauth4-1}
\bibliography{ref} 
%--------------------------------------------------------------------
\end{document}

%% file: appendix_1.tex
\appendix
\section{Linear-response scaling}\label{app:kappa}

For harmonics dominated by linear response, the final-state flow 
magnitude may be written approximately as~\cite{STAR:2019zaf, Liu:2018hjh}:
\begin{equation}
    v_n \simeq \kappa_n^{\rm L}\,\varepsilon_n ,~ \kappa_n^{\rm L} \propto \exp(-n^2\beta),
\end{equation}
where the response coefficient, $\kappa_n^{\rm L}$, contains the damping associated with the medium evolution. The $\beta$ encodes the strength of viscous attenuation. Under this approximation, combinations such as:
\begin{equation}
\zeta_{nm} = \frac{v_n^{1/n^2}}{v_m^{1/m^2}}, ~ \zeta_{nm}^{'} = \zeta_{nm} \frac{\varepsilon_m^{1/m^2}}{\varepsilon_n^{1/n^2}}
\end{equation}
largely cancel the leading acoustic damping factor.

\begin{figure}[th]
  \centering
  \includegraphics[width=1.0\linewidth]{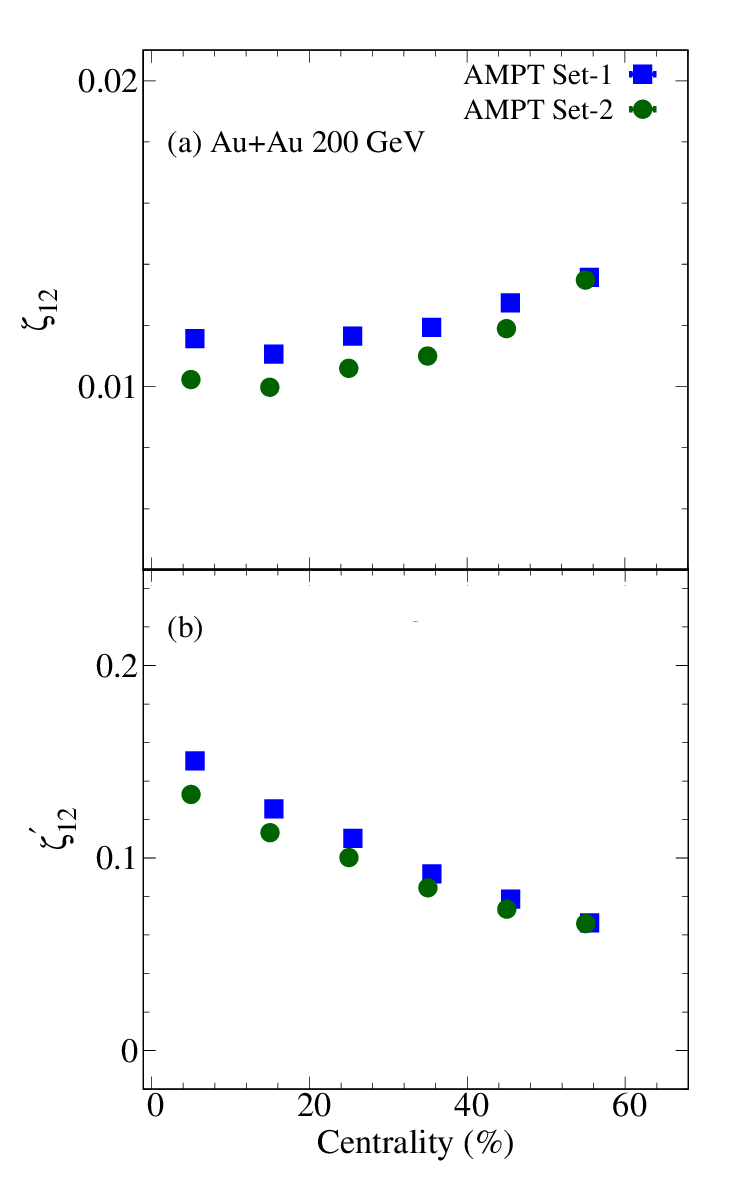}
  \caption{
  Centrality dependence of the $v_n$ ratios $\zeta_{12}$ (a) and the eccentricities scaled ratios $\zeta^{'}_{12}$ for the two AMPT settings used in this work.
  }
  \label{fig:app_v1_scaling1}
\end{figure}
\begin{figure}[th]
  \centering
  \includegraphics[width=1.0\linewidth]{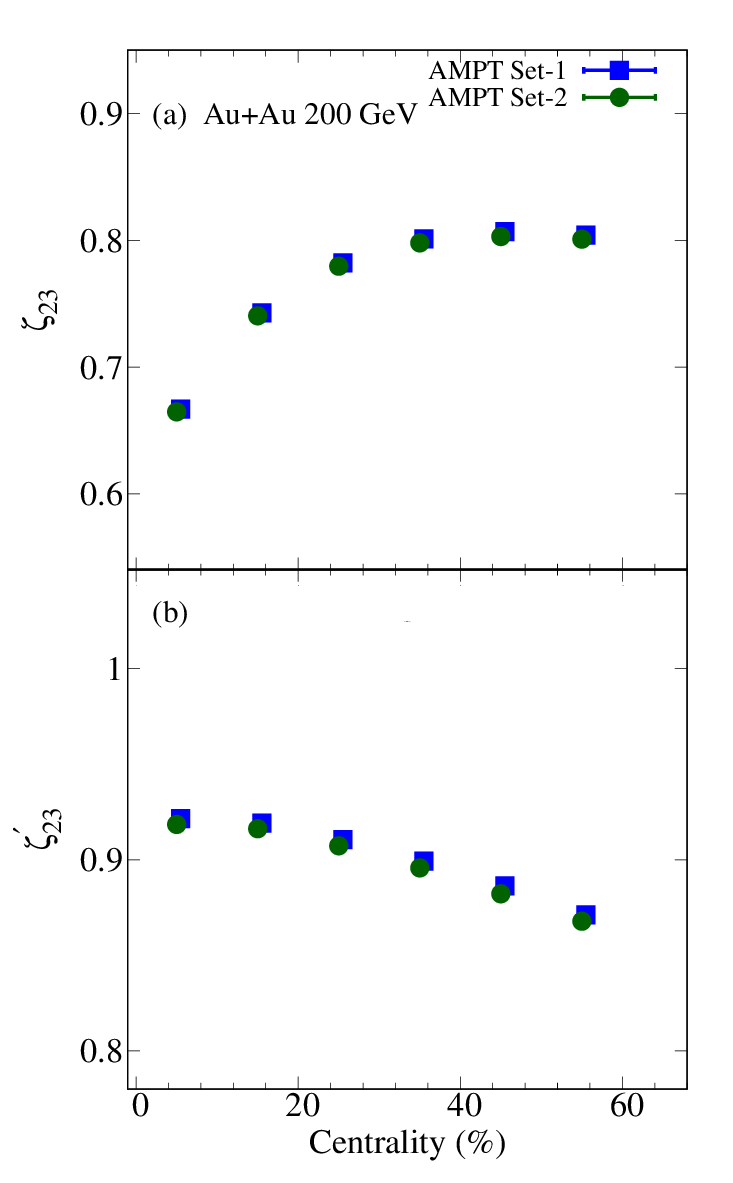}
  \caption{
  Centrality dependence of the $v_n$ ratios $\zeta_{23}$ (a) and the eccentricities scaled ratios $\zeta^{'}_{23}$ (b) for the two AMPT settings used in this work.
  }
  \label{fig:app_v1_scaling2}
\end{figure}

Figures~\ref{fig:app_v1_scaling1}~\ref{fig:app_v1_scaling2} shows that $\zeta_{23}$ and $\zeta^{'}_{23}$ exhibit no sensitivity to the AMPT transport setting, consistent with this approximate cancellation. By contrast, the ratios $\zeta_{12}$ and $\zeta^{'}_{12}$ retains a visible separation between the two AMPT settings. If the rapidity-even dipolar response followed the same simple acoustic scaling as $v_2$ and $v_3$, this ratio would be expected to show a stronger cancellation of the leading transport dependence. The remaining separation therefore suggests that the dipolar response is not fully captured by the same simple linear acoustic-scaling ansatz. A more quantitative separation of these effects requires dedicated studies and is left for future work.

\section{Comparison of event activity in AMPT and HIJING}\label{app:Comparison}

Because HIJING and string-melting AMPT employ different hadronization mechanisms, their final-state particle yields need not be identical at fixed centrality. The event activity in the three model samples used in this study is therefore compared.
\begin{figure}[th]
  \centering
  \includegraphics[width=0.99\linewidth]{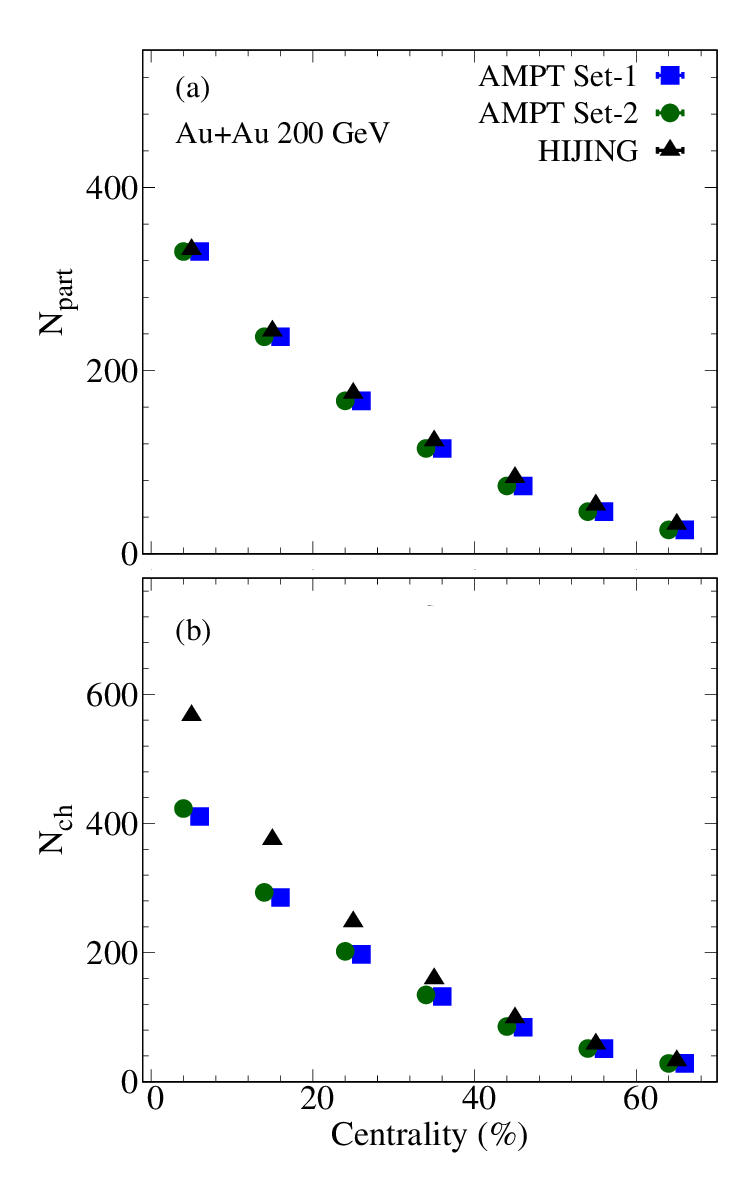}
  \caption{
Centrality dependence of (a) the mean number of participating nucleons, $N_{\rm part}$, and (b) the mean charged-particle multiplicity, $N_{\rm ch}$, in HIJING and in the two AMPT transport settings for Au+Au collisions at $\sqrt{s_{NN}}=200$~GeV. The charged-particle multiplicity is evaluated for the final-state charged hadrons satisfying $0.2<p_T<3.0$~GeV/$c$ and $|\eta|<0.5$.
  }
  \label{fig:ampt_hijing_mult}
\end{figure}

Figure~\ref{fig:ampt_hijing_mult} shows the centrality dependence of the mean number of participating nucleons, $N_{\rm part}$, and the mean charged-particle multiplicity, $N_{\rm ch}$, for HIJING and the two AMPT transport settings. The charged-particle multiplicity is evaluated
using final-state charged hadrons within
\[
0.2 < p_T < 3.0~{\rm GeV}/c,
\qquad
|\eta| < 0.5.
\]
The three samples have nearly identical $\langle N_{\rm part}\rangle$ values within each centrality interval, showing that the same geometric event classes are being compared.

A visible difference in $\langle N_{\rm ch}\rangle$ is observed in central collisions, where HIJING produces a larger charged-particle yield than either AMPT setting. This difference is expected because HIJING hadronizes through string fragmentation, whereas the string-melting version of AMPT employs partonic transport followed by quark coalescence and hadronic rescattering. The absolute multiplicity difference decreases toward peripheral collisions. In particular, the peripheral samples do not exhibit a large separation that could substantially complicate the use of HIJING as a control for momentum-conservation correlations.

This comparison is especially relevant because global momentum-conservation effects become increasingly important toward peripheral collisions as the event multiplicity decreases. In this
region, the HIJING and AMPT multiplicities are comparatively close, and the remaining model-to-model difference is modest relative to the large centrality evolution of the multiplicity itself. Consequently, the strong reduction of the HIJING first-harmonic correlations after application of the $\eta$-dependent weight is unlikely to be an artifact of an exceptionally large multiplicity mismatch in the peripheral region.